\documentclass{emulateapj}

\begin{document}


\title{A Troublesome Past: Chemodynamics of the Fornax dwarf spheroidal}


\author{N.C. Amorisco\altaffilmark{1} and N.W. Evans\altaffilmark{1}}
\affil{Institute of Astronomy, University of
  Cambridge, Madingley Road, Cambridge CB3 0HA, UK}
\email{amorisco@ast.cam.ac.uk, nwe@ast.cam.ac.uk}



\begin{abstract}
We present compelling evidence for the complexity of the Fornax dwarf
spheroidal. By disentangling three different stellar subpopulations
in its red giant branch, we are able to study in detail the dependence
between kinematics and metallicity. A well-defined ordering in
velocity dispersion, spatial concentration, and metallicity is evident
in the subpopulations. We also present evidence for a significant
misalignment between the angular momentum vectors of the old and
intermediate-age populations.  According to the HST measurement of
Fornax's proper motion, this corresponds to counter-rotation. These
ingredients are used to construct a novel evolutionary history of the
Fornax dwarf spheroidal, characterized as a late merger of a bound
pair.
\end{abstract}




\section{Introduction}

As the size and quality of datasets on the local population of dwarf
Spheroidals (dSphs) increase, evidence for the complexity of these
systems is intensifying. The observed variety of properties and
structures gives each dwarf its own distinctive character. Their star
formation histories, for instance, are well known to differ
significantly.  These often give rise to multiple generations of
stars, but the mechanisms that drive an intermittent star formation
must be investigated on a case-by-case basis.  Being able to untangle
the properties of such stellar subpopulations is crucial in advancing
our understanding of how they came in place.

After the Sagittarius dwarf, Fornax is the second most luminous
dSph orbiting the Milky Way and the only one (in addition to Sagittarius) to
posses an associated Globular Cluster system~\citep{vdB98}.  This
is not Fornax's only peculiarity: for example, significant asymmetries
in the isophotes were recorded as early as \citet{Ho61}, \citet{Esk88}
and then confirmed in \citet{Ir95}. Recent deep photometric surveys
\citep[][B06 in the following]{Ste98, Sav00, Bat06} have revealed a rich and
prolonged star formation, comprising old stars ($\gtrsim$ 10 Gyr), 
a dominant population of intermediate age stars, as
well as stars of only a few hundred Myr.

All systematic studies of Fornax's star formation history
\citep{Col08, Pino11, deB12} record a significant starburst at
approximately 4 Gyr ago. The trigger of this activity, though, is
still debated. On the one hand, at least three stellar overdensities
have been identified in Fornax \citep{Col04, Col05, deB12}, and
interpreted as reminiscent of the shell features observed in
elliptical galaxies \citep{Mal80}. In this picture, Fornax collided
with a low-mass sub-halo in the relatively recent past and swallowed
it. On the other hand, the details are not free from difficulties,
especially caused by the inferred metallicity and age of the stars
composing the innermost clump \citep{Ol06, Col08}. These are about 1.5
Gyr old, which makes the timing of the global starburst
problematic. They are also relatively metal-rich, which in turn does
not implicate an external origin for their gas. Furthermore, a recent
collision is quite unlikely when the energetics of encounters in a
virialized Milky Way halo is considered \citep{DeR04}.

After the pioneering work of~\citet{Wal11} (WP11 in the following), 
in this {\it Letter}, we provide new evidence for Fornax's complexity
by disentangling different stellar subpopulations in the red giant
branch (RGB) and by studying in detail their kinematics. Rather than a
division into two subpopulations \citep[][B06, WP11]{Sav00}, one into three
is preferred by the data. The identified subpopulations show
significant differences in line-of-sight (LOS) velocity dispersion and
higher order moments. Even more striking is the incompatibility
between their detected rotations, with the relative rotation vector
inclined $\sim 40^{\circ}$ with respect to the isophotal major axis.
Finally, we use these new ingredients to inform a novel formation
scenario for the Fornax dSph.


\section{Three Bayesian Subpopulations}

A step forward in modelling the superposition of chemodynamically
distinct stellar populations in dSphs has been achieved very recently
by WP11.  They devise a Bayesian technique that is able to
quantify and separate the distributions of projected radii, velocities
and metallicities of distinct subcomponents. Here, we use this
technique extensively. Given a spectroscopic dataset, each star is
associated to a set of probabilities of membership, one to each of the
identified stellar subpopulations. This is useful, as it allows us to
avoid any rigid color or metallicity cut, which inevitably degrades
the data. Instead, these probabilities can be used in deriving robust
and detailed kinematics for each subcomponent~\citep[e.g.,][A12 in the
  following]{Am12}. Each population is assumed to have a Plummer
surface density, a Gaussian metallicity distribution and a Gaussian
LOS velocity distribution (the latter hypothesis is relaxed in Section 4). Also, 
this technique is able to correct for the selection effects introduced by the
spatial distribution of the spectroscopic sample, hence all inferred quantities
-- such as half-light radii and fractions of giants -- are genuine global quantities.

WP11 apply this technique to the (approximately 2500) red giants in
the \citet{Wa09d} spectroscopic dataset (W09 in the following), and
look for a decomposition in, at most, two subcomponents.  Indeed, they
distinguish a dominant, colder and more concentrated metal-rich (MR)
stellar population, from a more diffused, kinematically hotter
metal-poor (MP) population. However, we reckon that such a division
into two subpopulations is not stable with respect to removal of the
kinematic information. If only the spatial distribution and the
metallicities of the giants are used (i.e. discarding the LOS
velocities), a different division is recovered. This is not the case
if the presence of a third population is allowed, as shown in Table~1.
In fact, a three-population division is a significantly better
description: since the two models are nested, the comparison is
straightforward and a simple maximum-likelihood ratio (MLR) is
sufficient. Following \citet{Ea71}, we find that despite the penalty
given by the additional free parameters, the probability of obtaining
such a MLR by pure chance is less than $3\times
10^{-6}$. Additionally, no evidence is found in the data to justify
the addition of a fourth population.

While the clear correlations between metallicity, spatial distribution
and velocity dispersion are naturally maintained in passing from two
to three subpopulations, the data prefer to accommodate separately a
small fraction of MR giants, which in turn have colder kinematics and
are more centrally concentrated. A division into three subpopulations
also best traces Fornax's star formation history, allowing at the same
time for stars that are older and younger than the bulk of
intermediate-metallicity (IM) giants. In this sense, the presence of
such a feature can be easily understood, as the abundant 1-2 Gyr old
population of stars, traced for example by the blue loop stars (B06 or
de Boer et al. 2012), are also expected to sneak into the RGB.

\begin{figure}
\centering
\includegraphics[width=\columnwidth]{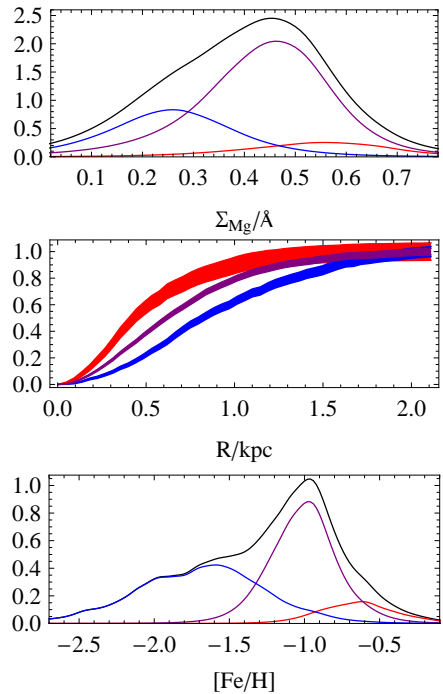}
\caption{Upper and middle panels: metallicity distributions and 
  cumulative mass distributions (one sigma errors) for the \citet{Wa09d} dataset. Lower panel: 
  metallicity distributions for the \citet{Bat06} dataset. 
  Color-coding is associated with increasing metallicity.}
\end{figure}

The top panel of Fig.~1 shows the final metallicity subdivision of the
W09 dataset. The MR peak is scarcely visible against the strong and
broad IM population, whilst the middle panel shows the cumulative 
mass distributions of the three populations, 
illustrating the importance of combining evidence
from the metallicity and the spatial distribution of the giants. A confirmation of
the statistical importance of the MR feature comes from the
independent spectroscopic dataset of B06. This has a different spatial
distribution and contains at least 400 giants that do not appear in the
W09 dataset. Furthermore, metallicities have been calibrated from
spectra in the Ca II triplet region, and are affected by smaller
uncertainties, resulting in a significantly less broadened metallicity distribution
in the lower panel of Fig.~1. 
We apply the same statistical technique and again
obtain that a division into three populations is a
significantly better description (the probability of obtaining
such a MLR by chance is $\lesssim 5\times 10^{-4}$). The MR feature sits at
$[$Fe/H$]\approx-0.65$, and is more concentrated than the other two
subpopulations\footnote{We do not record information on the fractions
  or scale radii of the three subpopulations as obtained from the B06
  dataset.  Unlike the metallicity, these are not reliable unless the
  spatial sampling bias is taken into account.}.

\begin{table*}
\begin{center}
\caption{Two and three population divisions of the Fornax dwarf}
\begin{tabular}{lccccc}
\tableline
\tableline
 Stellar & $\langle \Sigma_{\rm Mg}\rangle$ & StD$(\Sigma_{\rm
   Mg})$ & $R_{\rm h}$ & Fraction of &
 $\langle\sigma\rangle$ \\
Population& [in $\AA$] & [in $\AA$] & [in pc] & Population & [in kms$^{-1}$] \\
\tableline
Metal-poor & $0.31\pm0.025$  & $0.09\pm0.015$  & $860\pm60$  & $0.40\pm0.09$ & $14.4\pm0.6$ \\
     & $0.24\pm0.015$  & $0.04\pm0.015$  & $1050\pm90$ & $0.19\pm0.04$ & -            \\

Metal-rich & $0.47\pm0.01$   & $0.081\pm0.005$ & $556\pm24$  & $0.60\pm0.09$ & $10.0\pm0.6$ \\
     & $0.447\pm0.007$ & $0.091\pm0.004$ & $598\pm16$  & $0.81\pm0.04$  &  - \\
\tableline

Metal-poor & $0.27\pm0.018$  & $0.07\pm0.01$ & $888\pm50$ & $0.31\pm0.06$ & $14.3\pm0.6$ \\
     & $0.26\pm0.015$  & $0.05\pm0.01$ & $935\pm65$ & $0.27\pm0.09$ & -            \\
 
Intermediate & $0.46\pm0.01$   & $0.058\pm0.006$ & $605\pm30$ & $0.56\pm0.05$ & $11.3\pm0.7$ \\
     & $0.453\pm0.009$ & $0.065\pm0.008$ & $610\pm25$ & $0.63\pm0.07$ & - \\

Metal-rich & $0.53\pm0.035$ & $0.11\pm 0.02$ & $480\pm45$ & $0.13\pm0.04$ & $8.6\pm 1$ \\
     & $0.56\pm0.05$  & $0.10\pm 0.025$& $437\pm55$ & $0.10\pm0.04$ & - \\
\tableline
\end{tabular}
\tablecomments{Top-part: detailed results for a two-population division.
Lower part: a third population is allowed. Results are collected for the case in which the kinematic
information is retained (first line for each population), or discarded
(second line).}
\end{center}
\end{table*}

\section{A complex rotation pattern}

\begin{figure}
\centering
\includegraphics[width=.9\columnwidth]{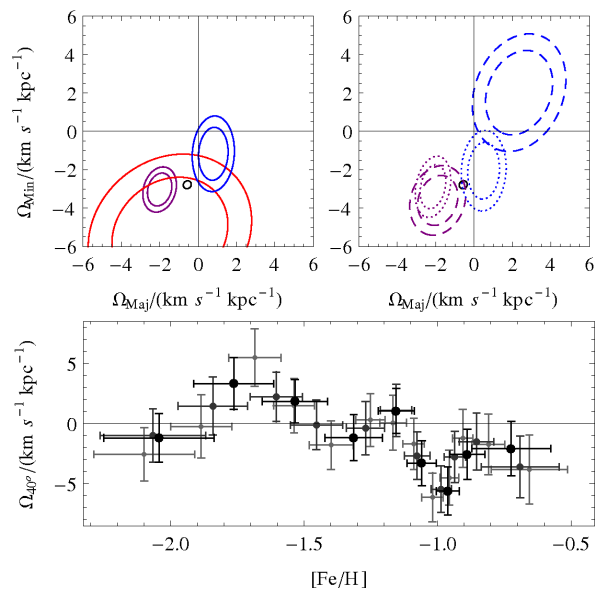}
\caption{Top-left: rotation pattern from the datasets of \citet{Wa09d}, 
 one and two sigma contours for each stellar population, color-coding as in Fig.~1; in black, Fornax's PM as
 from \citet{Pia07}. Top-right: rotation for inner (dotted ellipses) and outer (dashed ellipses) stars
 for the IM and MP population.
 Lower panel: the intrinsic rotation signal (after subtraction of the PM)
 along the direction of the relative rotation, as a function of
 metallicity, for the \citet{Bat06} dataset. Datapoints of growing sizes (and darker shades) are
 obtained from larger subsamples.}
\end{figure}

Since Fornax is an extended object, those stars that are misaligned
with the barycenter have their LOS velocities affected by a
geometrical projection of the systemic proper motion (PM)
\citep{Fea61}. This projection is equivalent to an apparent solid-body
rotation; hence, any rotation signal detected in the LOS velocities is
actually the superposition of any {\it intrinsic} and {\it apparent}
rotation:
\begin{equation}
\vec{\Omega}_{\rm tot} = \vec{\Omega}_{\rm int} + \vec{\Omega}_{\rm app}\ ,
\label{totrot}
\end{equation}
where $\vec{\Omega}_{\rm app}$ is proportional to the systemic PM
$\vec\mu$ (e.g. eqn (3) in A12). For clarity, our frequencies are 
defined so that
\begin{equation}
v_{\rm los}(x_{\rm Maj}, x_{\rm Min}) = 
 \left(\begin{array}{c}
{\Omega}_{\rm Maj}\\
{\Omega}_{\rm Min}
\end{array}
\right)  \cdot (x_{\rm Maj}, x_{\rm Min})\ ,
\label{freqdef}
\end{equation}
with all cartesian components referring to the major and minor axes of the isophotes.

\citet{Wal08} have recently measured the total rotation signal in
Fornax. They find it in agreement with the direct astrometric
determination of Fornax's PM, as obtained by \citet[][henceforth
  P07]{Pia07} using HST data. It was then natural to interpret this
result as evidence that Fornax has very little intrinsic rotation in
the LOS direction.  Indeed, both the W09 and the B06 datasets are
consistent with this interpretation, although this simple picture
remains valid only as long as all the giants in each dataset are
considered together. When we draw subsamples, we detect a strong
dependence of the total rotation signal on metallicity, with a
richness of detail that is surprising for such a small system.

In order to measure total LOS rotations $\vec{\Omega}_{\rm tot}$, we
adopt the same maximum-likelihood technique as in \citet{Wal08}. Our
probabilities of membership allow us to measure the rotational
properties of each subcomponent. As a compromise between size of the
subsamples and excessive contamination, we only consider giants with
at least 50\% chance of belonging to a specified population, but our
results are qualitatively stable against changes in this choice.
Results are shown in Fig.~2.

The measure we obtain for the MR feature has a large uncertainty, due
to the limited number of stars ($\approx80$ giants with $p_{\rm
  MR}>0.5$), as well their spatial concentration. On the other hand,
the rotations of the MP and the IM populations are precise enough to
clearly detect a {\it relative} rotation:
\begin{equation}
\vec{\Omega}_{\rm rel} \equiv \vec{\Omega}_{\rm tot}^{\rm IM} - \vec{\Omega}_{\rm tot}^{\rm MP}= \vec{\Omega}_{\rm int}^{\rm IM} - \vec{\Omega}_{\rm int}^{\rm MP}\ .
\label{relrot}
\end{equation}
  This is inclined at $\sim 40^{\circ}$ with respect to the major axis
  and provides a natural justification to the angular behaviour of the
  asymmetries detected in the LOS velocity distribution by A12 (see
  their Fig.~12). Its magnitude is independent of the systemic PM and
  is non-negligible, amounting to at least 3 kms$^{-1}$kpc$^{-1}$ --
  as a comparison, Fornax's tidal radius is $\sim 2.85$ kpc
  \citep{Ir95}. Furthermore, if we assume that the PM measured by P07
  is correct, then this relative rotation is also a {\it
    counter}-rotation.

The B06 dataset confirms this picture. In fact, since their
metallicites are more accurate, it provides an even more detailed
description, as shown in the lower panel of Fig.~2. A comparison with
Fig.~1 shows that peaks at positive and negative values of
$\Omega_{40^{\circ}}$ trace the peaks in the metallicity distribution,
respectively, of the MP and IM populations. Nonetheless, additional
structure is also visible.  The most significant lies in the MP end of
the distribution, where the rotational velocity $\Omega_{40^{\circ}}$
drops again at negative values. This feature is in fact rather
thought-provoking because it cannot be interpreted as an effect of
contamination of the MP population from the other
subpopulations. Still, it also appears in the W09 dataset. In fact,
additional indications that the MP population may indeed have a richer
structure come from the top-right panel of Fig.~2, in which the W09
subsamples pertaining to the IM and MP populations are split into two
circular annuli each. While for the IM the innermost giants (dashed
ellipses) share the same rotation as the outermost ones (dotted
ellipses), we detect a significantly more pronounced relative rotation
in the innermost MP giants (approximately within 1 kpc).


\section{Dispersions and Higher Order Moments}

\begin{figure}
\centering
\includegraphics[width=\columnwidth]{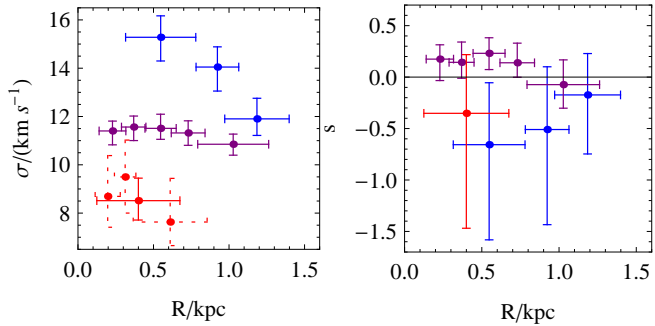}
\caption{Velocity dispersion and symmetric deviations in circular
  annuli for the three stellar subpopulations.}
\end{figure}

A12 have recently developed a Bayesian implementation for the
extraction of the shape information of line profiles from discrete
data. We apply that here and, for the first time, measure resolved
kinematics of the stellar subpopulations in Fornax. It is probably
worth mentioning that, in order to obtain reliable information on the
higher order moments, it is necessary to use the results on the
division into three populations as obtained by discarding the
kinematic information (see Section 2).

The left panel of Figure~3 collects the velocity dispersions profiles
-- in circular annuli -- for the three identified subpopulations, with
the same color-coding as in Fig.~1 (additional red-dotted points use a
smaller bin-size for the MR population). It is reassuring to find that
a Bayesian dissection only based on spatial positions and
metallicities also yields significantly different kinematic profiles.
In fact, also the shape of the velocity distributions shows significant
differences. We recall that the symmetry parameter $s$ is analogous to
the Gauss-Hermite coefficient $h_4$, and quantifies -- with respect to a Gaussian ($s=0$) -- 
how `peaked' ($s>0$) or `flat-topped' ($s<0$) the velocity distributions is (see A12 for further details).
A12 recorded a preference for a mild flat-toppedness in
Fornax ($s\approx-0.3$), which we can now interpret as being due,
at least in part, to the detected relative rotation. Instead, once isolated,
the IM population prefers quasi-Gaussian profiles, with a mild
tendency for positive values of $s$.

\section{A Merger of a Bound Pair?}

We have presented a detailed view of the rich kinematical properties
of the apparently puny Fornax dSph. Both spectroscopic datasets (B06
and W09) support a Bayesian division into three populations of its RGB, 
with a new MR feature being identified at
$[$Fe/H$]\approx-0.65$, containing approximately a tenth of the
giants. Both datasets also show a strong dependence between stellar
kinematics and metallicity. This includes a well-defined ordering of
the three populations in velocity dispersion and spatial
concentration, but it is most dramatic in a clear misalignment between
the angular momentum vectors of the MP and the IM populations. The
magnitude of this difference is independent of the systematic PM of
Fornax, but, if we assume that the measurement by P07 is correct, then
this misalignment also translates into a counter-rotation. The
populations are then rotating in opposite directions along an axis
inclined at $\sim 40^{\circ}$ with respect to the major axis.

Together with other evidence in the literature, the abundance of
features complicates the task of a coherent reconstruction of Fornax's
evolutionary history.  For instance, a merging scenario seems to be
naturally invoked by the shell features present in the photometry
\citep{Col04, Col05}. Even so, this is made less convincing by the
realization that the gas fueling the stars in such overdensities is
more likely from Fornax itself, given its significant pre-enrichment
\citep{Ol06, Col08}.  Furthermore, the merger scenario has to overcome
the difficulty of having a collision between two sub-haloes orbiting
the Milky Way at such late times \citep{DeR04}. On the other hand, it
is almost impossible to reconcile the misalignment between the angular
momenta of the MP and IM population without interactions. If we are
skeptical about the PM measurement by P07, we might relax the argument
for counter-rotation, and then invoke a tidal origin for the
different rotational properties of the MP stellar population.
Nonetheless, the fact that it is actually the innermost MP stars that
have a faster relative rotation rather than the outermost ones argues
against this, since those stars sit at the same distance from the center
as most of the IM stars.

Furthermore, supporting evidence is provided by the fact that the spatial
distribution of the youngest stars in Fornax seems to have knowledge
of the direction of the relative rotation. \citet{Ste98} note that the main
sequence stars ``display a clearly flattened distribution on the sky,
with a long axis in the east-west direction'' (see their Fig.~12), hence
in perfect alignment with $\vec{\Omega}_{\rm rel}$. It is then appealing to advocate that this central
structure, composed by stars that are younger than a few 100 Myr and
whose geometry is uncorrelated with Fornax's isophotes, is reminiscent
of an elongated bar. If so, its spatial alignment is also the
direction of its intrinsic rotational velocity. In turn, this is
aligned with the intrinsic rotations of the MP and IM populations only
if the PM is not significantly different from P07.

We suggest that a solution for these inconsistences is that Fornax is
a merger of a bound pair. First, if Fornax and its perturber have some
relative orbital motion at early times, say with frequency
$\vec\Omega_{\rm orb}$, then interactions with the old MP population
provides an efficient means of heating and preferentially stripping
the progradely rotating stars \citep{Don10, Rea06}.  As a result,
whatever the initial internal orbital structure, the MP population is
left with a counter-rotating signature (with respect to
$\vec\Omega_{\rm orb}$).  Second, the nearby presence of a companion
increases significantly the chances of retaining the gas expelled by
the first generations of stars in Fornax. Its escape is made more
difficult by the increased total mass and by the gas-dynamics of the
interactions with the companion. As dynamical friction eventually
causes the decay of the perturber's orbit, the infalling gas naturally
has both a high metallicity and a prograde rotational signature, hence
imprinted in the younger stellar populations.  This process may be
prolonged, and the localized overdensities~\citep{Col04} and the
central bar of main sequence stars may represent the last stages
before the supply of gas is extinguished.

How massive might such a putative companion be?  By assuming that the
luminosity in the RGB is about one half of the total in Fornax, that
the luminosity functions of the three sub-populations are the same,
and that the metal poorest tail contains about one third of the MP
giants, we get $L \approx 7\times 10^5 L_{\odot}$, roughly comparable
to Canes Venatici I, Draco, Sextans, Ursa Minor, although this may
just be a lower limit. These dSphs have masses of $\approx 4\times
10^7 M_{\odot}$ within 1.7 half light radii \citep[e.g.,][]{Am11},
therefore probably more than a few $10^8 M_{\odot}$ in total. While
this is still too low to make a late collision probable, it is instead
enough to significantly perturb the progradely rotating stars in
Fornax, and to ensure an infall timescale of a fraction of the Hubble
time. We conclude that this picture collects the benefits
of the merging scenario while overcoming most of its limits.




\acknowledgments NA thanks STFC and the Isaac Newton Trust for
financial support.  We thank Adriano Agnello, Giuseppe Bertin, Elena
D'Onghia, Jorge Pe{\~n}arrubia, Justin Read and Matthew Walker for
inspiring conversations and useful criticism, as well as the referee,
Mario Mateo, for helping us to improve the manuscript.


\begin{thebibliography}{}

\bibitem[Amorisco \& Evans(2011)]{Am11} Amorisco, N.~C., \& Evans,
  N.~W.\ 2011, \mnras, 411, 2118

\bibitem[Amorisco \& Evans(2012)]{Am12} Amorisco, N.~C., Evans,
  N.~W.\ 2012, MNRAS, in press, arXiv:1204.5181 (A12)

\bibitem[Battaglia et al.(2006)]{Bat06} Battaglia, G., Tolstoy, E.,
  Helmi, A., et al.\ 2006, \aap, 459, 423 (B06)

\bibitem[Coleman et al.(2004)]{Col04} Coleman, M., Da Costa, G.~S.,
  Bland-Hawthorn, J., et al.\ 2004, \aj, 127, 832

\bibitem[Coleman et al.(2005)]{Col05} Coleman, M.~G., Da Costa, G.~S.,
  Bland-Hawthorn, J., Freeman, K.~C.\ 2005, \aj, 129, 1443

\bibitem[Coleman \& de Jong(2008)]{Col08} Coleman, M.~G., de Jong,
  J.~T.~A.\ 2008, \apj, 685, 933

\bibitem[de Boer et al. (2012)]{deB12} de Boer, T., 2012, PhD thesis,
  University of Groningen

\bibitem[Del Pino et al.(2011)]{Pino11} Del Pino, A., Aparicio, A.,
  Gallart, C., Hidalgo, S.\ 2011, EAS Publications Series, 48, 77

\bibitem[De Rijcke et al.(2004)]{DeR04} De Rijcke, S., Dejonghe, H.,
  Zeilinger, W.~W., Hau, G.~K.~T.\ 2004, \aap, 426, 53

\bibitem[D'Onghia et al.(2010)]{Don10} D'Onghia, E., Vogelsberger, M.,
  Faucher-Giguere, C.-A., Hernquist, L.\ 2010, \apj, 725, 353

\bibitem[Eadie et al.(1971)]{Ea71} Eadie, W.~T., Drijard, D., James,
  F.~E., Roos, M., Sadoulet, B. \ 1971, Statistical Methods in
  Experimental PhysicsAmsterdam: North-Holland, 1971,

\bibitem[Eskridge(1988)]{Esk88} Eskridge, P.~B.\ 1988, \aj, 
96, 1614 

\bibitem[Feast et al.(1961)]{Fea61} Feast, M.~W., Thackeray, A.~D.,
  Wesselink, A.~J.\ 1961, \mnras, 122, 433

\bibitem[Hodge(1961)]{Ho61} Hodge, P.~W.\ 1961, \aj, 66, 249 

\bibitem[Irwin \& Hatzidimitriou(1995)]{Ir95} Irwin, M.,
  Hatzidimitriou, D.\ 1995, \mnras, 277, 1354



\bibitem[Malin \& Carter(1980)]{Mal80} Malin, D.~F., Carter, D.\ 1980,
  \nat, 285, 643

\bibitem[Olszewski et al.(2006)]{Ol06} Olszewski, E.~W., 
Mateo, M., Harris, J., et al.\ 2006, \aj, 131, 912 


\bibitem[Piatek et al.(2007)]{Pia07} Piatek, S., Pryor, C., Bristow,
  P., et al.\ 2007, \aj, 133, 818 (P07)

\bibitem[Read et al.(2006)]{Rea06} Read, J.~I., Wilkinson, 
M.~I., Evans, N.~W., Gilmore, G., \& Kleyna, J.~T.\ 2006, \mnras, 366, 429 

\bibitem[Saviane et al.(2000)]{Sav00} Saviane, I., Held, E.~V.,
  Bertelli, G.\ 2000, \aap, 355, 56

\bibitem[Stetson et al.(1998)]{Ste98} Stetson, P.~B., Hesser, J.~E.,
  Smecker-Hane, T.~A.\ 1998, \pasp, 110, 533

\bibitem[van den Bergh(1998)]{vdB98} van den Bergh, S.\ 1998, 
\apjl, 505, L127 

\bibitem[Walker et al.(2008)]{Wal08} Walker, M.~G., Mateo, 
M., Olszewski, E.~W.\ 2008, \apjl, 688, L75 

\bibitem[Walker et al.(2009)]{Wa09d} Walker, M.~G., Mateo, M.,
  Olszewski, E.~W.\ 2009, \aj, 137, 3100 (W09)

\bibitem[Walker \& Pe{\~n}arrubia(2011)]{Wal11} Walker, M.~G.,
  Pe{\~n}arrubia, J.\ 2011, \apj, 742, 20 (WP11)


\end{thebibliography}
\end{document}